# Tough sheets of nanowires produced floating in the gas phase†

Richard S. Schäufele [ab], Miguel Vazquez-Pufleau [a] and Juan J. Vilatela *[a]

[a] IMDEA Materials, Madrid, 28049, Spain. E-mail: juanjose.vilatela@imdea.org; Tel: +34 915493422

[b] Department of Applied Physics, Universidad Autónoma de Madrid, Cantoblanco, 28049, Madrid, Spain

## Abstract

Assembling nanostructured building blocks into network materials unlocks macroscopic properties inaccessible with monolithic solids, notably toughness and tolerance to electrochemical alloying. A method is reported for large-scale, continuous synthesis of silicon nanowires (SiNWs) suspended in the gas phase and their direct assembly into macroscopic sheets. Performing gas-phase growth of SiNWs through floating catalyst chemical vapor deposition using an aerosol of gold nanoparticles eliminates the need for substrates, increasing the growth rate by a factor of 500, reaching 1.4 µm s$^{-1}$ and leading to very long SiNWs. The combined high aspect ratio (>210) and large concentration of SiNWs in the gas-phase (1.5 × 10$^7$ cm$^{-3}$) enable the formation of macroscopic solids solely composed of percolated SiNWs, such as free-standing sheets and continuous metre-long SiNW tapes. Sheet samples of small diameter SiNWs (<25 nm) combine extraordinary flexibility in bending, tensile ductility around 3%, and over 50-fold higher toughness than Si-based anodes (fracture energy 0.18 ± 0.1 J g$^{-1}$). This synthesis and assembly process should be applicable to virtually any one-dimensional inorganic nanomaterial producible by thermochemical methods.

## New Concepts

We present a new universal route to produce continuous sheets of inorganic nanowires directly assembled through their growth suspended in a gas stream, with textile-like properties and an order-of-magnitude higher toughness than monolithic analogues. Drawing on evidence that chemical vapour deposition can be conducted with catalyst aerosols, the manuscript demonstrates the synthesis of silicon nanowires (SiNWs) floating in the gas phase, at high particle concentration and unprecedented fast growth rate, enabling the direct formation of percolating networks and fabrication of continuous macroscopic solids solely composed of SiNWs. Their network structure of bundled nanowires makes them flexible in bending and tolerant to knotting, similar to fabrics and in stark contrast with monolithic ceramics. Deformation through network reorganisation and nanowire stress transfer in shear lead to high tensile toughness, with exceptionally high fracture energy compared to regular Si materials, and similar to sheets of high-performance nanocarbons or boron nitride nanotubes. The synthesis method should be applicable to virtually any one-dimensional inorganic nanomaterial producible by thermochemical methods. Moreover, controlling percolation through aspect ratio and concentration in the gas phase gives access to novel network materials ranging from transparent conductors to dense fibres using the same process; with an inherent toughness attractive for mechanically augmented devices.

# 1 Introduction

One-dimensional (1D) inorganic nanostructures are fascinating objects that combine quantized optoelectronic properties, enormous surface-to-volume ratio and often an extremely high degree of crystallinity. As an archetypal example, silicon nanowires (SiNWs) of small diameter show visible photoluminescence,[1] high photocatalytic activity,[2] and tolerance to large expansions under electrochemical conversion reactions,[3] all distinct from the properties of bulk silicon. Exploiting the properties of 1D nanostructures requires integrating a large number of them into macroscopic ensembles. 1D inorganic nanostructures can be assembled into a range of architectures,[4] but most often bound to a rigid supporting surface, reminiscent of a substrate-growth process or because the macroscopic ensemble lacks mechanical integrity. Some exceptions include growth of ultra-light silicon carbide aerogels using sacrificial templates,[5] plasma-assisted synthesis of boron nitride nanotube (BNNT) felt,[6] and spinning of fibres of carbon nanotubes (CNTs).[7–10] In the latter case, controlled growth of long carbon nanotubes and their direct assembly from the gas phase into aligned fibres[7] has led to mechanical tensile strength and toughness superior to almost any engineering material,[11] combined with higher thermal conductivity than copper[12] and the flexibility of staple yarns, amongst other properties. A fascinating prospect is the development of a generic route to assemble 1D inorganic nanostructures into continuous networks forming free-standing macroscopic materials. Considering that most inorganic 1D nanostructures can be synthesized by substrate-based chemical vapor deposition (CVD),[13] it is promising to eliminate the substrate by using an aerosol of catalyst nanoparticles floating in a gas stream[14] to grow nanowires in the gas phase and perform their assembly into macroscopic networks in one stage. In this work, we demonstrate large-scale, continuous synthesis of SiNWs in the gas-phase by floating catalyst chemical vapour deposition (FCCVD) and direct assembly into macroscopic structures resembling tough fabrics or sheets. When composed of small diameter SiNWs, the sheets are flexible, can be knotted, and undergo mechanical deformation as a textile, absorbing on average 0.18 J g$^{-1}$ under tensile deformation, more than an order of magnitude above monolithic Si wafers used in optoelectronics or Si microparticle battery electrodes.

# 2 Results and discussion

We performed the synthesis of SiNWs through chemical vapour deposition of silane (SiH$_4$) using an aerosol of Au nanoparticles to enable SiNW growth unsupported and suspended in the gas-phase. The synthesis reaction is carried out in a vertical tube furnace with entry ports at the top for gases and Au nanoparticles generated by thermal evaporation, and a chamber at the bottom to collect the SiNW material (Fig. 1a). Incoming silane decomposes at the floating catalyst nanoparticles as they travel through the length of the reactor, initially forming the Au–Si eutectic and then, upon supersaturation, leading to the growth of thin nanowires preferentially in the ⟨110⟩ direction at the end of the catalyst nanoparticles.[15] Growth of SiNWs by floating catalyst chemical vapour deposition eliminates the need for a substrate to support the catalyst, thus enabling a continuous synthesis process (see Supporting Video, ESI†) and direct assembly into different macroscopic formats. Fig. 1b shows an example of a continuous tape of SiNWs of metre-scale and Fig. 1c a photograph of porous solid solely made of SiNWs collected from the gas phase, including electron micrographs of the constituent nanowires at different scales.

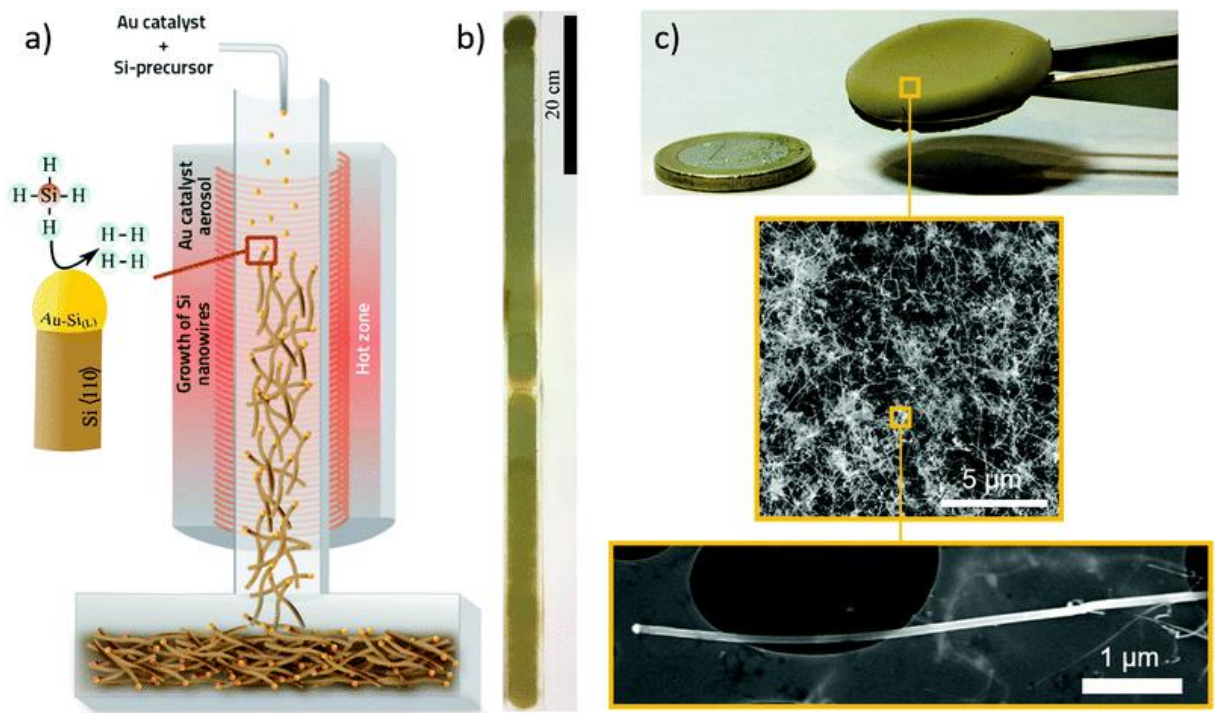

**Fig. 1** Continuous synthesis of macroscopic networks of SiNWs by floating catalyst chemical vapour deposition. (a) Schematic of the synthesis process based on catalytic decomposition of $SiH_4$ in the presence of an aerosol of Au particles. (b) Photograph of continous SiNW tape on the metre scale. (c) Free-standing SiNW material and electron micrographs of the porous ensemble of long SiNWs.

We have found a wide window of reaction conditions leading to the successful growth of SiNWs, but optimum sample quality in terms of crystallinity, reduced content of self-nucleated Si nanoparticles and low tapering is currently obtained by conducting the reaction at 650 °C, using silane at a concentration of 1.0% in a mixture of $N_2$ and $H_2$ as carrier gases. A typical sample produced under these conditions has over 95 vol% of SiNW, with around 1 vol% Au residual catalyst. The SiNWs are long, straight and mono-crystalline (diamond structure), with their main axis parallel to ⟨110⟩. Examples of electron micrographs are shown in Fig. 2. We determined the distributions of SiNW diameter and aspect ratio by image analysis of electron micrographs (Fig. 2 and Fig. S3, ESI†). They show an average diameter of $\Phi = 20.0 \pm 5.5$ nm and an aspect ratio of $s = 214.3 \pm 63.8$. There is no observed correlation between nanowire aspect ratio and diameter; it then follows that thicker nanowires tend to be longer (reaching up to 13 µm) (Fig. S4, ESI†). This suggests that the availability of $SiH_4$ is not the reaction rate-limiting step and that growth rate depends on the catalyst particle size. To avoid overestimating the length in case electron micrograph analysis is biased towards thicker SiNWs, the average length can be calculated from the product of average diameter and average aspect ratio, which results in $L = 4.3 ± 2.3$ μm.

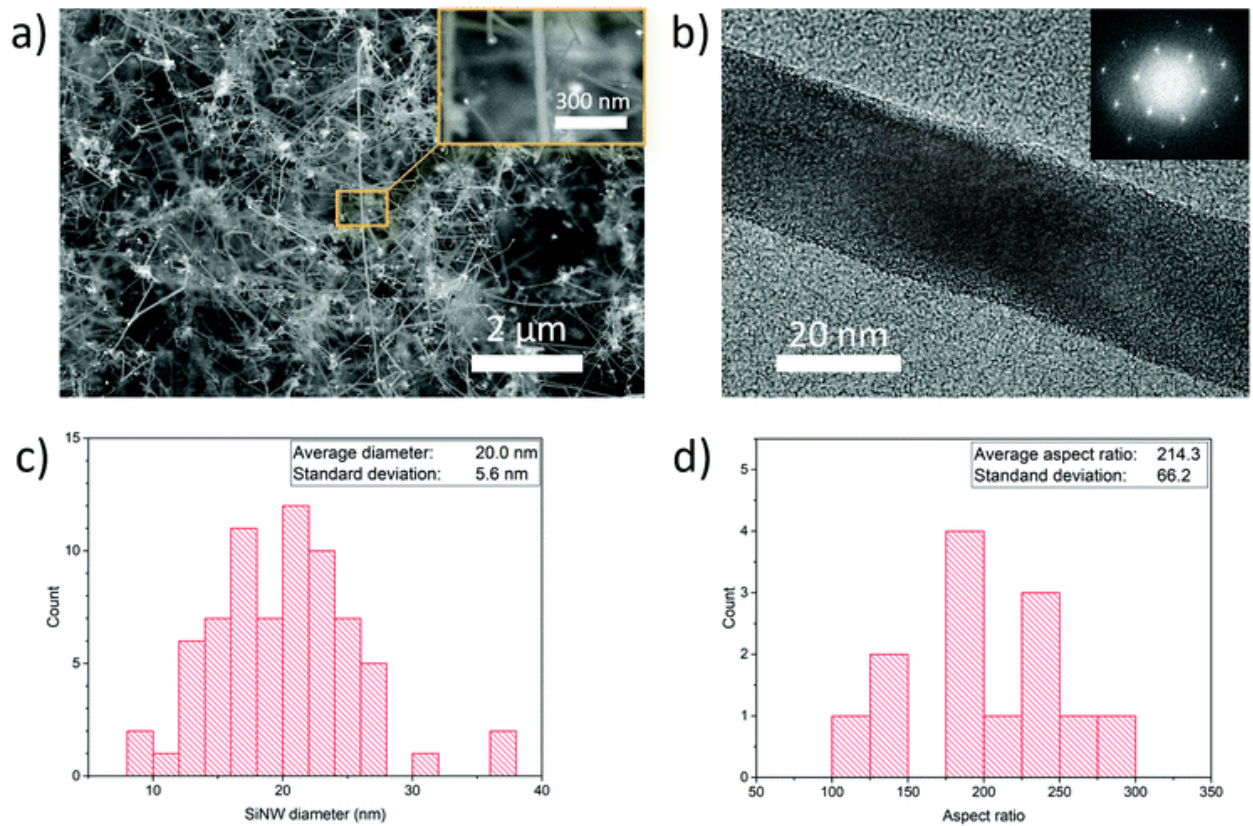

**Fig. 2** Size distribution of SiNWs produced in the gas-phase. (a) Electron micrograph of a network of long SiNWs. (b) HRTEM and FFT inset of monocrystalline SiNWs with the growth direction ⟨110⟩. (c) Diameter and (d) aspect ratio distributions of SiNWs determined from electron mirographs. The product of average diameter and aspect ratio gives an average length of $L = 4.3 ± 2.3$ μm.

A key aspect of the reaction enabling the formation of macroscopic networks of SiNWs lies in the simultaneous achievement of a very fast growth rate and high throughput. For a typical residence time of 8 s, a high estimate calculated from gas velocity through the reactor hot zone, the average growth rate can reach 1400 nm s$^{-1}$, which is orders of magnitude higher than conventional gold-catalysed substrate growth of SiNWs, typically at around 2–15 nm s$^{-1}$ (Fig. S1, ESI†). Similar enhancements in growth rate compared to substrate-based processes have been observed in CNTs[7] and GaAs NWs[14] grown by FCCVD, and thus seem to be a generic feature of this synthesis mode. We attribute them to faster kinetics for precursor transport in floating catalyst conditions resulting from a much larger separation between NWs, which avoids reactant depletion zones, and the fact that both the precursors and the catalyst can diffuse simultaneously. But very importantly, the growth of long, highly crystalline SiNWs reported here is achieved at high

1D nanoparticle concentration in the reactor. Based on throughput determined from direct measurement of sheet sample mass, we estimate a typical nanowire number concentration of $1.5 \times 10^7$ cm$^{-3}$ in the gas phase.

The combined high aspect ratio and high concentration in the reactor favour the aggregation of SiNWs through the formation of entanglements and contact between SiNWs, thus enabling collection into mechanically robust macroscopic ensembles. A useful process descriptor can be derived from the general theory of particle networks: a continuous (percolating) particle network forms above a critical volume concentration ($v_p$), determined by the particle excluded volume and thus related to its aspect ratio ($s$) as $v_p \approx 0.5/s$.[21] Drawing from this simple relation, we use the product of aspect ratio and particle volume concentration in the reactor ($vs$) as a figure of merit to describe the probability for 1D objects produced in the gas-phase to form continuous networks. Expressed in terms of particle diameter ($\Phi$), length ($L$) and number concentration ($n$) it leads to $vs = 0.25\pi\Phi nL^2$. This simple "aerogelation parameter" captures, for example, the importance of increasing concentration and length simultaneously. In Table 1, we compare the values of $vs$ for various processes of gas-phase synthesis of nanomaterials leading to different macroscopic formats. The synthesis conditions used in this work give a value of $4.3 \times 10^{-6}$. This is orders of magnitude higher than the existing methods for gas-phase synthesis of thin films of 1D nanomaterials, such as CNT transparent conductors ($6.3 \times 10^{-9}$)[17] or GaAsNW networks for optoelectronics ($4 \times 10^{-8}$),[14] purposely kept very dilute to control optical absorption and maintain transparency. The present values of $vs$ are however, currently lower than processes for the *in situ* formation of fibres, either of CNTs grown by FCCVD ($2.5 \times 10^{-2}$)[16] or of BNNTs produced by plasma-assisted methods ($5.3 \times 10^{-2}$).[20] Under synthesis conditions explored so far, we visually observed macroscopic aggregates of SiNWs in the gas phase at the exit of the reactor (Fig. S2, ESI†), but they are discontinuous. Nevertheless, current values of $vs$ are sufficient to produce integrated engineering materials exclusively made up of SiNWs, such as sheets, and determine their macroscopic properties, as discussed below.

**Table 1** Production parameters for FCCVD synthesis of 1D nanomaterials in different formats

| Nanomaterial type | Particle number concentration (# cm$^{-3}$) | Average aspect ratio | Aerogelation parameter | Throughput (g/day) | Ensemble format |
|---|---|---|---|---|---|
| SiNW (this work) | $1.5 \times 10^7$ | 214 | $4.3 \times 10^{-6}$ | 0.2 | sheets |
| CNT[16] | $1.1 \times 10^9$ | 33000 | $2.5 \times 10^{-2}$ | 2.33 | fibres |
| CNT[17] | $5 \times 10^5$ | 4000 | $6.3 \times 10^{-9}$ | $\approx 4.8 \times 10^{-6}$ | thin film |
| GaAs[14] | $1 \times 10^6$ | 20 | $4 \times 10^{-8}$ | $\approx 0.024$ | thin film |
| BNNT[18,19] | $\approx 7.8 \times 10^8$ | 31 | $1.6 \times 10^{-7}$ | $\approx 0.48$ | powder |
| BNNT[20] | $4.2 \times 10^{10}$ | 5000 | $5.3 \times 10^{-2}$ | 840 | fibres |

Free-standing macroscopic samples can be easily produced by collecting the SiNWs on a filter during gas-phase synthesis, and then manually removing the sample from the filter. They resemble thin fabrics in terms of their overall toughness during handling. As a result of their network structure of building blocks with a diameter much smaller than the sample thickness, as in a regular textile fabric, they have large flexibility in bending. In Fig. 3a, we show a strip of SiNW sheet with an overhand knot, equivalent to withstanding deformations at a radius of curvature of below 0.8 mm. As a further indication of mechanical robustness, centimetre-long samples as thin as 3 µm are free-standing.

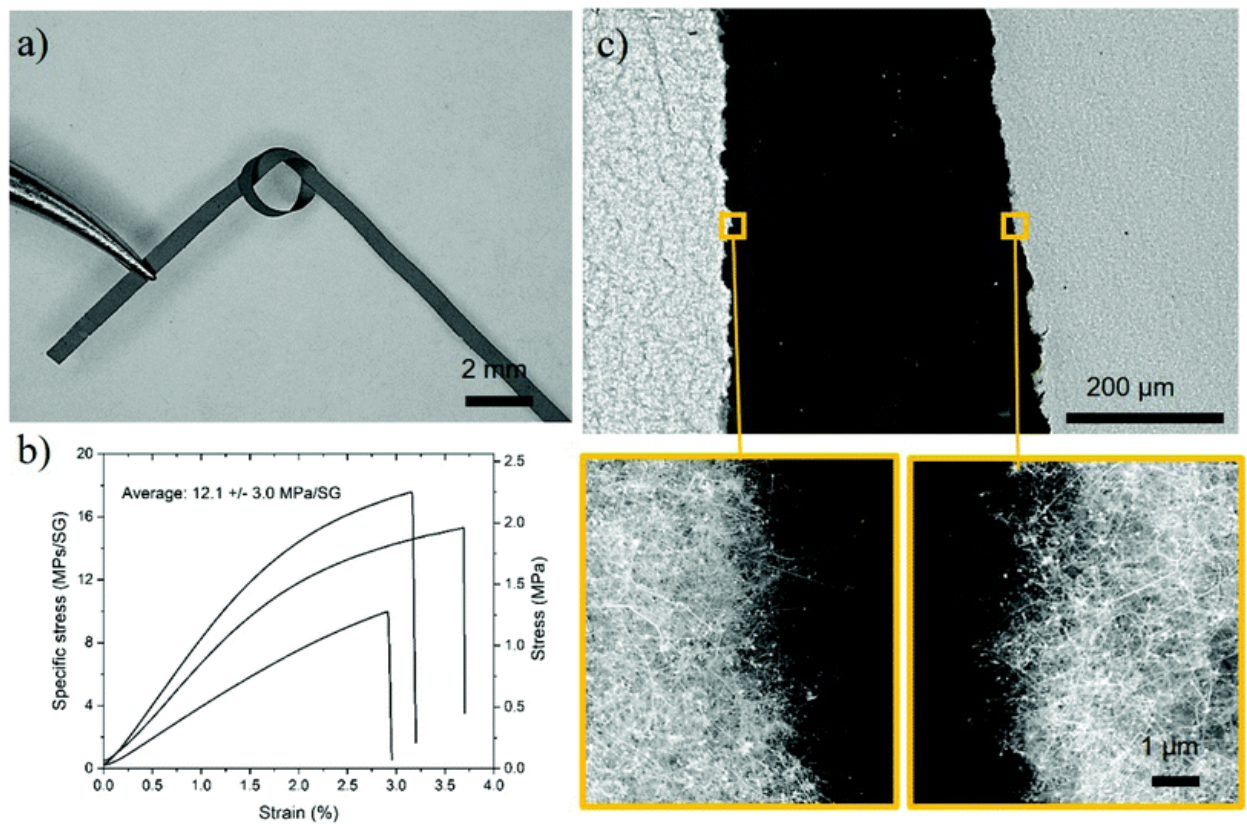

**Fig. 3** Mechanical properties of SiNW sheets and similarity to textiles. (a) Photograph of a sample with an overhand knot, demonstrating high flexiblity in bending. (b) Example of a stress–strain curve with high fracture energy from elasto-plastic deformation and correspondingly high ductility. (c) Micrographs showing ragged fracture surfaces resulting from SiNW slippage, similar to a staple fibre fabric. SG: specific gravity.

We have also performed tensile tests on rectangular sheets, with a focus on samples composed of SiNWs of a relatively small diameter of 20.0 ± 5.6 nm, and a length of 4.2 ± 2.3 µm, controlled through adjustment of the size distribution of Au catalyst particles and the precursor concentration (see Experimental section). Smaller NW diameters were purposely targeted to favour their compaction. By virtue of their smaller bending stiffness (scaling as $\Phi^4$), these SiNWs behave as flexible rods that can form domains of close NW contact (Fig. S5, ESI†), similar to bundles, increasing stress transfer in shear between them.[22]

Fig. 3b presents examples of specific stress against engineering strain (use of specific units avoids uncertainty with the determination of the cross-section, which is problematic in porous samples). The first aspect that stands out is their large ductility around 3%. This is far higher than the strain-to-break of monolithic inorganic materials, such as bulk silicon (<0.1%), for example. The sloping profile of the stress–strain curve is indicative of a network material with elasto-plastic tensile behaviour combining stress development through elastic deformation of the nanowires, entanglements and frictional stress from nanowire slippage. Indeed, the fracture surface is ragged, and resembles that of a textile or staple fibre fabric like cotton (Fig. 3c). The frictional stress developed during plastic deformation provides the material with high overall toughness, with an average density-normalized fracture energy of 0.18 ± 0.1 J $g^{-1}$ and as high as 0.36 J $g^{-1}$. This fracture energy is significantly above both bulk Si (0.014 J $g^{-1}$) and regular Si microparticle battery anodes (0.003 J $g^{-1}$),[23] and comparable to recently-reported CNTs/Si composite battery anodes with nearly theoretical capacity and extended cyclability enabled by a fracture energy of 0.21 J $g^{-1}$.[23]

Finally, we compared different materials with a network structure made up of nanoscale building blocks. Fig. 4 shows a plot of the density-normalized tensile fracture energy against the ratio of sample density over its theoretical bulk density. This representation enables visualisation of the build-up of toughness with increasing compaction of 1D elements in the network. At the low end are materials with ultra-low density such as $SiO_2$ aerogels composed of a rigid cellular structure of thick nanofibres (>200 nm), which behave as foams that can withstand transverse compaction but have otherwise low tensile toughness, with a fracture energy of around 0.01 J $g^{-1}$. When of a smaller diameter (<50 nm), 1D nanostructures can compact more easily and form larger interfacial areas of contact, thus forming tougher materials resembling fabrics that are capable of reaching higher tensile fracture energies of 0.1–1 J $g^{-1}$. The SiNW sheet samples produced by FCCVD in this work are in this category, as well as sheets of BNNTs[6] or of CNTs,[27] often referred to as buckypapers. In these sheets, the 1D nanostructures are predominantly aligned along the main axis but randomly oriented in the plane. Increased density and load-bearing capacity requires a higher packing and alignment of building blocks in the longitudinal direction, as in aligned CNT fibres reaching tensile fracture energies of 10–100 J $g^{-1}$.

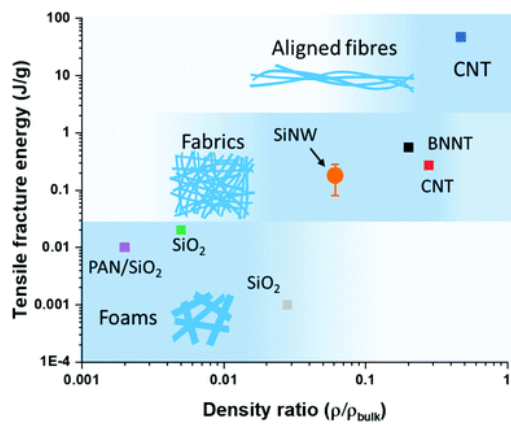

**Fig. 4** Tensile fracture energy and density ratio of SiNW sheets and other porous materials with network structures. Tensile fracture energy builds up from contact between elements in the network and is related to their compaction. Large diameter building blocks are rigid and lead to brittle foams,[24–26] small diameter building blocks enable packing into sheets with moderate properties,[6,27] and when highly aligned into high-performance fibres.[11]

Producing macroscopic ensembles of SiNWs that cover the range from foams to aligned fibres should be within reach, if SiNW diameter and length are controlled during growth in the gas phase. Combining this control with the possibility to translate the present FCCVD method to the enormous library of 1D nanomaterials that can be synthesized through conventional CVD unlocks a strategy to produce a myriad of new macroscopic engineering materials from 1D nanobuilding blocks. The inherent toughness of network structures makes them particularly attractive for energy storage[28] and conversion applications,[29] especially those relying on integration into complex shapes for use in mechanically-augmented devices, such as in wearable electronics[30,31] and energy-storing structural composites.[32]

Another interesting aspect of the direct conversion of gas precursors to fully integrated solids is the elimination of processing steps requiring handling of the nanomaterials in powder form. As powders, nanomaterials are generally more prone to individualisation, often requiring special safety handling measures.[33] Although in most envisaged applications, nanowires are contained by some form of encapsulation, toughness resulting from strong interparticle forces and entanglements inherently minimises exposure to individualised particles during mechanical manipulation. Nevertheless, a precautionary approach should be applied to SiNWs, even in light of *in vivo* studies indicating early rapid SiNW clearance after being intratracheally instilled in rats[34] and promising applications when interfaced with living cells.[35]

## 3 Conclusions

In conclusion, we show continuous synthesis of SiNW networks in the gas-phase and direct association into macroscopic solids. Conducting SiNW growth using an aerosol of catalyst nanoparticles increases the growth rate by orders of magnitude compared to substrate-based CVD, attributed to faster kinetics for precursor transport in floating catalyst conditions by avoiding reactant depletion zones and enabling diffusion of both the precursors and catalyst. Critically, the growth of long building blocks is achieved at high particle concentrations in the reactor, thus favouring the formation of percolated networks of SiNWs and association into macroscopic solids. An implication of this work is the demonstration of a generic route for the growth of ultralong 1D inorganic nanostructures in the gas phase and their direct assembly into macroscopic solids. This strategy should be applicable to a large number of 1D nanostructures producible through substrate-based CVD. Controlled organisation into porous solids seems particularly promising for energy storage and conversion applications.

The product of aspect ratio and volumetric concentration ($vs$) is introduced as a descriptor to guide improvements towards the fabrication of high-performance materials made up of 1D nanostructures. Large increases in the aspect ratio may be attainable by temperature-gradient control[36] or using alternative Si precursors,[37] previously shown to produce SiNWs of 1–20 mm grown on substrates. Related gas-phase synthesis processes for ultrahigh aspect ratio CNTs[8,16] or large throughput in plasma-based growth of BNNTs,[20] suggest promising avenues for the simultaneous improvement of both aspect ratio and volumetric concentration. Increases in SiNW length and reaction conversion will also translate into a large cost drop from the corresponding reduction of Au concentration in the final material. Sheets of SiNWs of small diameter, long SiNWs resemble textiles and with orders of magnitude higher toughness than ceramic materials and Si microparticle electrodes. This augurs high capacity retention as lithium ion battery electrodes, for example.

# 4 Experimental

**Synthesis of SiNWs by FCCVD**

The FCCVD reactor consists of a chamber for catalyst nanoparticle generation, a reaction tube and a collection chamber. An aerosol of gold particles was generated *via* resistive heating of a gold source at approximately 1500 °C. A nitrogen gas flow was introduced at the heated gold source with a continuous flow rate of 2 slm. Just before the reactor tube (stainless steel tube, diameter of 10 cm), silane gas and hydrogen gas were added. The typical nitrogen, hydrogen and silane ratio used is 41 : 58 : 1. The temperature of the reactor was set to 650 °C. As SiNWs were grown in the reaction tube, the structures were collected down-stream *via* filtration using a conventional porous paper. These synthesis conditions led to an approximate throughput of 0.2 g per day, based on the mass of collected material. Samples of thicker NWs and lower purity could also be produced at a throughput of 1 g per day. Typical sample weight was around 20 mg.

SiNWs and residual Au nanoparticles were calculated from image analysis of SEM micrographs using brightness, contrast and shape differences to produce areal maps of filtered composition.

## SiNW diameter and length determination

SiNW diameter and aspect ratio distributions were obtained by image analysis of scanning electron micrographs at high magnification (Fig. S3, ESI†). Length was calculated from the product of diameter and aspect ratio.

## Sample preparation

Preparation of free-standing samples consisted in depositing SiNWs on a vacuum filter at the end of the FCCVD reactor, typically for a collection time of 30 minutes. The material was then mechanically removed from the filter and then densified with isopropanol. This densification procedure was applied to all samples shown in the manuscript except those in Fig. 1 and 2. By weighing the sheets we could easily determine the areal density. Samples for tensile tests were cut out from the sheet using a scalpel.

## Tensile tests

Tensile tests were performed with a Textechno Favimat tensile tester, at a strain rate of 10% min$^{-1}$. Sample dimensions were determined from optical micrographs of each sample. The typical width was 0.6 mm, and the thickness was 25 μm. Volumetric density could then be determined from areal density and thickness. Discarding specimens that broke at the grips, 36 samples were tested in total, 23 at a gauge length of 5 mm, 5 at 2 mm, and 8 at 1 mm. No significant difference in tensile strength was found at smaller gauge lengths. Data were corrected for machine compliance, obtained from tensile tests on commercial poly-aramid fibres. Stress–strain curves in the main manuscript are for 5 mm gauge-length samples. Tensile fracture energy values for the main paper were calculated from the data in Table S1 (ESI†). For SiNWs, data are calculated from the 10 best measurements with clear evidence of a genuine fracture not induced by grips of defects introduced during manipulation. Density ratios are calculated assuming a maximum density corresponding to hexagonal closed packed bundles of solid rods each with the theoretical bulk density of the material.

## Conflicts of interest

There are no conflicts to declare.

## Acknowledgements

The authors are grateful for the generous financial support provided by the European Union Seventh Framework Program under grant agreement 678565 (ERC-STEM) and by MINECO (RyC-2014-15115).

## Notes and references

1. J. Holmes, K. Johnston, R. Doty and B. Korgel, *Science*, 2000, **287**, 1471–1473.
2. M. Shao, L. Cheng, X. Zhang, D. D. D. Ma and S.-T. Lee, *J. Am. Chem. Soc.*, 2009, **131**, 17738–17739.
3. C. Chan, H. Peng, G. Liu, K. McIlwrath, X. Zhang, R. Huggins and Y. Cui, *Nat. Nanotechnol.*, 2008, **3**, 31–35.
4. R. K. Joshi and J. J. Schneider, *Chem. Soc. Rev.*, 2012, **41**, 5285–5312.
5. L. Su, H. Wang, M. Niu, X. Fan, M. Ma, Z. Shi and S.-W. Guo, *ACS Nano*, 2018, **12**, 3103–3111.
6. P. Nautiyal, C. Zhang, A. Loganathan, B. Boesl and A. Agarwal, *ACS Appl. Nano Mater.*, 2019, **2**, 4402–4416.
7. Y.-L. Li, I. A. Kinloch and A. H. Windle, *Science*, 2004, **304**(5668), 276–278.
8. L. Ericson, H. Fan, H. Peng, V. Davis, W. Zhou, J. Sulpizio, Y. Wang, R. Booker, J. Vavro, C. Guthy, A. Parra-Vasquez, M. Kim, S. Ramesh, R. Saini, C. Kittrell, G. Lavin, H. Schmidt, W. Adams, W. Billups, M. Pasquali, W.-F. Hwang, R. Hauge, J. Fischer and R. Smalley, *Science*, 2004, **305**, 1447–1450.
9. B. Vigolo, A. Penicaud, C. Coulon, C. Sauder, R. Pailler, C. Journet, P. Bernier and P. Poulin, *Science*, 2000, **290**, 1331–1334.
10. M. Zhang, K. Atkinson and R. Baughman, *Science*, 2004, **306**, 1358–1361.
11. K. Koziol, J. J. Vilatela, A. Moisala, M. Motta, P. Cunniff, M. Sennett and A. Windle, *Science*, 2007, **318**, 1892–1895.
12. T. S. Gspann, S. M. Juckes, J. F. Niven, M. B. Johnson, J. A. Elliott, M. A. White and A. H. Windle, *Carbon*, 2017, **114**, 160–168.
13. L. Güniat, P. Caroff and A. Fontcuberta i Morral, *Chem. Rev.*, 2019, **119**, 8958–8971.
14. M. Heurlin, D. Lindgren, K. Deppert, L. Samuelson, M. Magnusson, M. Ek and R. Wallenberg, *Nature*, 2012, **492**, 90–94.
15. V. Schmidt, J. V. Wittemann and U. Gösele, *Chem. Rev.*, 2010, **110**, 361–388.
16. V. Reguero, B. Alemán, B. Mas and J. J. Vilatela, *Chem. Mater.*, 2014, **26**, 3550–3557.
17. K. Mustonen, PhD thesis, Department of Applied Physics, Aalto University School of Science, 2015.
18. M. J. Kim, S. Chatterjee, S. M. Kim, E. A. Stach, M. G. Bradley, M. J. Pender, L. G. Sneddon and B. Maruyama, *Nano Lett.*, 2008, **8**, 3298–3302.
19. S. Chatterjee, M. J. Kim, D. N. Zakharov, S. M. Kim, E. A. Stach, B. Maruyama and L. G. Sneddon, *Chem. Mater.*, 2012, **24**, 2872–2879.
20. A. Fathalizadeh, T. Pham, W. Mickelson and A. Zettl, *Nano Lett.*, 2014, **14**, 4881–4886.
21. I. Balberg, C. Anderson, S. Alexander and N. Wagner, *Phys. Rev. B: Condens. Matter Mater. Phys.*, 1984, **30**, 3933–3943.
22. J. J. Vilatela, J. A. Elliott and A. H. Windle, *ACS Nano*, 2011, **5**, 1921–1927.
23. S.-H. Park, P. King, R. Tian, C. Boland, J. Coelho, C. Zhang, P. McBean, N. McEvoy, M. Kremer, D. Daly, J. Coleman and V. Nicolosi, *Nat. Energy*, 2019, **4**, 560–567.

24. Y. Si, X. Wang, L. Dou, J. Yu and B. Ding, *Sci. Adv.*, 2018, **4**, eaas8925.
25. Y. Si, J. Yu, X. Tang, J. Ge and B. Ding, *Nat. Commun.*, 2014, **5**, 5802.
26. J. C. Wong, H. Kaymak, S. Brunner and M. M. Koebel, *Microporous Mesoporous Mater.*, 2014, **183**, 23–29.
27. J. N. Coleman, W. J. Blau, A. B. Dalton, E. Muñoz, S. Collins, B. G. Kim, J. Razal, M. Selvidge, G. Vieiro and R. H. Baughman, *Appl. Phys. Lett.*, 2003, **82**, 1682–1684.
28. A. M. Chockla, J. T. Harris, V. A. Akhavan, T. D. Bogart, V. C. Holmberg, C. Steinhagen, C. B. Mullins, K. J. Stevenson and B. A. Korgel, *J. Am. Chem. Soc.*, 2011, **133**, 20914–20921.
29. K.-Q. Peng and S.-T. Lee, *Adv. Mater.*, 2011, **23**, 198–215.
30. W. A. D. M. Jayathilaka, K. Qi, Y. Qin, A. Chinnappan, W. Serrano-García, C. Baskar, H. Wang, J. He, S. Cui, S. W. Thomas and S. Ramakrishna, *Adv. Mater.*, 2019, **31**, 1805921.
31. B.-C. Zhang, H. Wang, Y. Zhao, F. Li, X.-M. Ou, B.-Q. Sun and X.-H. Zhang, *Nanoscale*, 2016, **8**(4), 2123–2128.
32. C. González, J. Vilatela, J. Molina-Aldareguía, C. Lopes and J. LLorca, *Prog. Mater. Sci.*, 2017, **89**, 194–251.
33. S. Sharifi, S. Behzadi, S. Laurent, M. Laird Forrest, P. Stroeve and M. Mahmoudi, *Chem. Soc. Rev.*, 2012, **41**, 2323–2343.
34. J. Roberts, R. Mercer, R. Chapman, G. Cohen, S. Bangsaruntip, D. Schwegler-Berry, J. Scabilloni, V. Castranova, J. Antonini and S. Leonard, *J. Nanomater.*, 2012, **2012**, 398302.
35. D. Nagesha, M. Whitehead and J. Coffer, *Adv. Mater.*, 2005, **17**, 921–924.
36. B.-C. Zhang, H. Wang, L. He, C.-J. Zheng, J.-S. Jie, Y. Lifshitz, S.-T. Lee and X.-H. Zhang, *Nano Lett.*, 2017, **17**, 7323–7329.
37. W. I. Park, G. Zheng, X. Jiang, B. Tian and C. M. Lieber, *Nano Lett.*, 2008, **8**, 3004–3009.

**Footnote**

† Electronic supplementary information (ESI) available: Experimental methods, comparison of SiNW growth rates for this work and substrate-based processes in the literature, evidence of macroscopic SiNW aggregates within the gas phase, example of an SEM micrograph for measuring SiNW diameters as well as data on aspect ratio, versus, SiNW diameter and average length calculation, evidence of bundling, details on the preparation of SiNW sheets and of mechanical data used for comparison, and a video showing continuous fabrication of SiNW sheets. See DOI: 10.1039/d0mh00777c